\documentclass[a4paper]{jpconf}
\newcommand{\psim}{\lower.5ex\hbox{$\; \buildrel \propto \over\sim \;$}}
\newcommand{\lesssim}{\lower.5ex\hbox{$\; \buildrel < \over\sim \;$}}
\newcommand{\gtrsim}{\lower.5ex\hbox{$\; \buildrel > \over\sim \;$}}
\usepackage{graphicx}
\begin{document}
\title{On leptonic models for blazars in the Fermi era}

\author{Charles Dermer$^1$ \& Benoit Lott$^2$}

\address{$^1$Code 7653, Space Science Division, U.S.\ Naval Research Laboratory, 4555 Overlook Ave., SW, Washington, DC 20375-5352}
\address{$^2$Centre d'\'Etudes Nucl\'eaires Bordeaux Gradignan, Universit\'e de Bordeaux, CNRS/IN2P3, UMR 5797, Gradignan, 33175, France}

\ead{charles.dermer@nrl.navy.mil}

\begin{abstract}
Some questions raised by {\it Fermi}-LAT data about blazars are summarized, along with attempts at solutions within the context of leptonic models. These include both spectral and statistical questions, including the origin of the GeV breaks in low-synchrotron peaked blazars, the location of the gamma-ray emission sites, the correlations in the spectral energy distributions with luminosity, and the difficulty of synchrotron/SSC models to fit the spectra of some TeV blazars.
\end{abstract}

\section{Introduction}
The 11 June 2008 launch of the Gamma-ray Large Area Space Telescope, subsequently renamed the Fermi Gamma ray Space Telescope, opened a new era into studies of radio galaxies and blazars. The Large Area Telescope (LAT) on Fermi, with a peak on-axis effective of $\approx 8000$ cm$^2$ and a single photon angular resolution for 68\% containment at 1 GeV of $\approx 0.8^\circ$ \cite{2009ApJ...697.1071A}, represents an improvement by 1 -- 2 orders of magnitude in sensitivity over its predecessors EGRET on the Compton Gamma-ray Observatory and AGILE \cite{2009A&A...502..995T}. In the Third EGRET Catalog of Gamma Ray Sources \cite{1999ApJS..123...79H}, 66 high-confidence ($\gtrsim 5\sigma$) detections of blazars and one probable radio galaxy, Centaurus A were reported. Ten high-latitude sources at high galactic latitude ($|b|>10^\circ$) were detected with significance $>10\sigma$. Now, just over three years into the Fermi mission, three important lists of {\it Fermi}-LAT AGNs have been or are about to be published:
\begin{enumerate}
  \item 
The LAT Bright AGN Sample (LBAS) \cite{2009ApJ...700..597A}. Based on the first three months of scientific operations of the Fermi satellite, the LBAS consists of 106 high-latitude sources with $TS\gtrsim 100$, or significance $\gtrsim 10\sigma$. It consists of 58 flat spectrum radio quasars (FSRQs), 42 BL Lac objects, 2 radio galaxies, namely, Centaurus A and NGC 1275 (3C 84), and 4 blazars with
unknown classification.

\item The First LAT AGN catalog (1LAC) \cite{2010ApJ...715..429A}. This catalog contains 671 sources that are associated at high confidence with  709 AGNs (some $\gamma$-ray sources have multiple associations). The associations are based on counterpart catalogs, such as the flat-spectrum 8.4 GHz CRATES (Combined Radio All-Sky Targeted Eight GHz Survey) Catalog \cite{2007ApJS..171...61H} and the Roma BZCAT blazar catalog \cite{2009A&A...495..691M}. The 1LAC clean sample is a subset of the high-confidence associations, and consists of 599 AGNs with no multiple associations or other analysis flags. These sources subdivide into 275 BL Lac objects, 248 FSRQs, 50 blazars of unknown type (either because a source lacks an optical spectrum, or the optical spectrum is inadequate to determine if it is a BL Lac or FSRQ), and 26 other AGNs. These ``other" AGNs include 7 radio galaxies/misaligned AGNs, 4 radio-loud narrow line Seyfert 1 (RL-NLSy1) galaxies, 4 narrow line radio galaxies, 3 starburst galaxies (which of course are not actually AGNs), and 9 other sources.

\item The Second LAT AGN catalog (2LAC) \cite{2LAC}. The 2LAC contains 1016 sources that are associated at high confidence with AGNs.  The 2LAC clean sample consists of 885 sources (46\% larger than the 1LAC clean sample), consisting of 395 BL~Lacs, 310 FSRQs, 156 blazars of unknown type, 8 misaligned AGNs, 4 RL-NLSy1 galaxies, 10 AGNs of other types, and 2 starburst galaxies (NGC 4945 has fallen out of the list).   

\end{enumerate}

Important questions that are impacted by or arise from the {\it Fermi}-LAT observations of blazars and radio galaxies, and  interpretation that relates to these questions within the context of leptonic models, will occupy the remainder of this short review.

\section{Important questions}

After summarizing the standard leptonic model for blazars, we consider some important questions raised by the {\it Fermi}-LAT data,  including
\begin{enumerate}
  \item Limitations of one-zone synchrotron/SSC model for TeV blazars
  \item Relation between aligned and misaligned jet sources
  \item Validity and meaning of the blazar sequence
  \item GeV spectral cutoffs of LSP and ISP blazars
  \item Super-Eddington luminosities of blazars
\end{enumerate}

\subsection{Leptonic Models for blazar SEDs}

The general outlines of blazar leptonic models are now well established for the standard one-zone model within the context of blazar unification \cite{1995PASP..107..803U}. In this paradigm, the basic ingredients of a blazar are an accretion disk, regions of broad and narrow emission-line gas, and a dusty IR torus. Radio-loudness is a consequence of collimated relativistic outflows of plasma directed along the polar axis of the accretion disk.  When observing at small angles to the plane of the accretion disk, the torus gas conceals the  broad-line region formed close to the black hole, so that only narrow emission lines are detected \cite{1993ARA&A..31..473A}. These are the Type II Seyfert galaxies or, if radio-loud, narrow line radio galaxies. Observations at larger angles to the plane of the accretion disk, where the torus no longer obscures the central region, reveal the broad emission lines. These are the Type I Seyferts or, if radio-loud, the broad-line radio galaxies. 

When viewed at very small angles with respect to the jet axis, the Doppler-boosted nonthermal radiation overwhelms the accretion-disk radiation and makes the familiar two-humped broadband blazar continuum \cite{2007Ap&SS.309...95B}. The featureless continuum  is made by nonthermal particles accelerated at shocks formed, for example, by colliding shells \cite{2010ApJ...711..445B} or through sheared flows in structured jets \cite{2000ApJ...528L..85A}. In the standard blazar model for flaring activity, a compact plasma zone filled with nonthermal electrons and positrons radiate synchrotron photons that form the highly polarized radio continuum that extends to optical frequencies in flat spectrum radio quasars (FSRQs), and to X-ray frequencies in the high-synchrotron peaked (HSP; with peak synchrotron frequency $\nu^{pk}_{syn}> 10^{15}$ Hz) BL Lac objects. Accompanying the synchrotron emission component is the synchrotron self-Compton (SSC) $\gamma$-ray component. This simple one-zone model successfully fits the SEDs of HSP BL Lac objects like Mrk 421 \cite{2011ApJ...736..131A} or Mrk 501 \cite{2011ApJ...727..129A}. The synchrotron/SSC model for these objects implies Doppler factors $\delta_{\rm D} \sim 10$ -- 50, comoving magnetic fields $B^\prime \sim 0.01$ -- 0.1 G, and absolute powers $\approx 10^{44}$ erg s$^{-1}$. The outflow Lorentz factors are compatible with values determined by superluminal radio observations \cite{2001ApJS..134..181J}, and the inferred jet powers are far below the Eddington luminosity for black holes with mass $\approx 10^{9} M_\odot$ that are thought to power these objects. 

The simple one-zone synchrotron/SSC model has difficulty to explain the SEDs of FRSQs and intermediate (ISP; $10^{14}$ Hz $< \nu^{pk}_{syn}< 10^{15}$ Hz)  BL Lacs like BL Lac itself \cite{2011ApJ...730..101A}, which have dominant $\gamma$-ray components that would require, for a synchrotron/SSC model, parameters far from equipartition, which would consequently place large demands on jet power.  External radiation fields, either from the disk \cite{2002ApJ...575..667D}, the broad-line region (BLR), or the IR torus \cite{2009ApJ...704...38S}, are then invoked to provide seed photons that are Compton-scattered to form an additional $\gamma$-ray emission component. Within this framework, precise detailed fits are made to sources like 3C 279 \cite{2007Ap&SS.309...95B}. Blazar modeling seemed somewhat satisfactorily explained at the close of the EGRET era with synchrotron/SSC models for X-ray selected BL Lac objects, and with the addition external Compton scattering components for FSRQs and radio-selected BL Lac objects.  

\subsection{Difficulties of simple blazar models}

Even before the launch of {\it Fermi}, cracks in our understanding of the blazar phenomenon became apparent from observations with TeV telescopes. The July 2006 flaring data of PKS 2155-305 observed with HESS \cite{2007ApJ...664L..71A} and Swift required $\delta_{\rm D} \gtrsim 100$ for  good synchrotron/SSC model fits \cite{2008ApJ...686..181F}. Very high energy (VHE; $\gtrsim 100$ GeV) emission from the FSRQ 3C 279 observed with the MAGIC telescope \cite{2008Sci...320.1752M} was found to pose severe difficulties to either synchrotron/SSC or external Compton scattering models \cite{2009ApJ...703.1168B}. Unexpectedly low extragalactic background light (EBL) intensities at the level of that implied by galaxy counts were required to make sense of the unusually hard deabsorbed TeV spectrum of  H 2356-309  and 1ES 1101-232 \cite{2006Natur.440.1018A}. Furthermore, doubts about the $\gamma$-ray emission site being located within the BLR were raised by  radio/$\gamma$-ray correlations \cite{2003ApJ...590...95L} and multwavelength observations of BL Lacertae, suggesting emission in an inner acceleration zone and at an outer recollimation zone  \cite{2008Natur.452..966M}.

These difficulties have been compounded with the advent of Fermi and long baseline GeV data correlated with targeted campaigns at radio, optical, X-ray, and VHE energies. Broadband modeling \cite{2010MNRAS.401.1570T} of a large sample of TeV BL Lacs found that most are well fit with a synchrotron/SSC model with $B^\prime \sim 0.01$ -- 1 G and $\delta_{\rm D} \sim 20$ -- 50, but with a second group of blazars displaying unusually low magnetic fields and large break electron Lorentz factors. These latter sources include 1ES 1101-232, 1ES 0229+200, and 1ES 0347-121, and are peculiar for BL Lac objects by being very weakly variable or nonvariable, and displaying very weak GeV fluxes.  They are now being used to make inferences about the level of the intergalactic magnetic field (IGMF) \cite{2010Sci...328...73N}, insofar as the absorbed primary TeV radiation  make a large flux of secondary electrons and positrons from attenuated $\gamma$ rays. These pairs would make GeV radiation by Compton scattering photons of the CMBR and would overproduce the measured Fermi flux unless the pairs are deflected by a sufficiently strong intergalactic magnetic field. The implied IGMF ranges from $\approx 10^{-18}$ G to $\approx 10^{-15}$ G \cite{2011ApJ...727L...4D,2011ApJ...733L..21D,2011MNRAS.414.3566T} depending on the IGMF correlation length and duration of the blazar engine.

If the IGMF is $\approx 10^{-16}$ -- $10^{-18}$ G, then a separate, quiescent GeV emission component will be present in the SED of HSP BL Lac objects. Indeed, this could account for the slowly varying GeV flux in sources like Mrk 501 \cite{2011arXiv1104.2801N}. The situation is further complicated if TeV blazars accelerate ultra-high energy cosmic rays (UHECRs) and produce a separate high-energy radiation component due to photopair production by UHECR protons while traveling through intergalactic space \cite{2010APh....33...81E,2010PhRvL.104n1102E}. This could affect inferences of the EBL based on the assumption that the highest energy photons are made at the source. This ambiguity can be resolved by observations of multi-TeV radiation by CTA and HAWC from sources like 1ES 0229+200, which can be used to discriminate between a photon-induced and UHECR proton-induced origin of the high-energy radiation and determine whether TeV blazars are sites of UHECR acceleration \cite{2011arXiv1107.5576M}.  
 
For FSRQs such as PKS 1510-089 \cite{2010ApJ...710L.126M} and 3C279 \cite{2010Natur.463..919A}, coherent optical polarization position-angle swings over timescales of many weeks with late-time flares coinciding with strong $\gamma$-ray activity seem to be  most simply explained if the emission is formed in a helical jet on multi-pc scales. It is not clear, however, how this relates to indications in PKS 1510-089 that the $\gamma$-ray flux precedes the optical flux by $\approx 13$ d  \cite{2010ApJ...721.1425A}.  Moreover, the interpretation of distant emission regions may collide with the inference about sharp GeV cutoffs in the SEDs of these sources, as described in the next section.

\subsection{GeV spectral cutoffs of LSP and ISP blazars}

Already by the time of the LBAS and the first paper on 3C 454.3 \cite{2009ApJ...699..817A} it was evident that the $\gamma$-ray spectra of FSRQs, which are nearly all low synchrotron-peaked (LSP) sources with $\nu F_\nu$ peak synchrotron frequency $\nu^{syn}_{pk}< 10^{14}$ Hz, as well as ISP objects such as AO 0235+164, display remarkably sharp cutoffs between 1 and 10 GeV. This feature is particularly evident in the compilation of spectra of Fermi bright blazars \cite{2010ApJ...710.1271A}, where a distribution of source-frame break energies $E^*_{br}$ as a function of apparent isotropic $\gamma$-ray luminosity $L_\gamma$ is displayed. Essentially all sources have $E^*_{br}$ within 3$\sigma$ of 5 GeV. Two other important aspects of this result should be mentioned: (1) The break energy changes only slightly  with flux state. In the case of 3C 454.3,  $E^*_{br}$ increases by  a factor of 2 when the flux varies by more than a factor of 20, as indicated by spectral analysis of the bright GeV blazar 3C 454.3  \cite{2010ApJ...721.1383A}. (2) The change in spectral index below and above the break is typically larger than 1 unit, ruling out naive radiative cooling models that predict a spectral break of 0.5 units.
 
This has generated considerable theoretical interest. A hybrid scattering scenario \cite{2010ApJ...714L.303F} involving both accretion-disk and BLR photons can reproduce the SED of 3C454.3, with the added feature that the spectral break energy is insensitive to location if the BLR has a wind-like profile, because the energy densities of both the accretion-disk and BLR radiation fields then scale with the same dependence on radius.  Poutanen and Stern argue in print  \cite{2010ApJ...717L.118P,2011arXiv1105.2762S} and at this conference that He II recombination line and continuum radiation attenuate a continuum GeV emission component through $\gamma\gamma$ pair production. For He II recombination radiation at energy $\approx 4\times 13.6 = 54.4$ eV or $\epsilon_s \cong 54.4/511000 = 1.06\times 10^{-4}$ (in $m_ec^2$ units), photons with energy $m_ec^2 \epsilon_1 > 2/\epsilon \cong 9.6$ GeV, close to $E^*_{br}$, can be strongly absorbed. The $\gamma$-ray emitting region must be found deep within the BLR for this model to work, and the excessive target photon energy density may introduce difficulties to fit the multi-band spectral data of 3C 454.3. 

A further possibility is that the breaks are due to Klein-Nishina effects on Compton-scattered Ly $\alpha$ photons \cite{2010ApJ...721.1383A}. GALEX observations of 3C 454.3, as reported by Bonnoli and coworkers \cite{2011MNRAS.410..368B}, reveal the presence of a strong Ly $\alpha$ radiation field in the BLR of 3C 454.3, amounting to a line luminosity of (2 -- 4)$\times 10^{45}$ erg s$^{-1}$. The onset of the Klein-Nishina decline takes place when $4\gamma\epsilon \cong 1$ for electrons with Lorentz factor $\gamma$ scattering Ly $\alpha$ target photons with dimensionless energy $\epsilon = 10.2/511000 = 2\times 10^{-5}$. Photons are Compton-scattered to energies  $\epsilon_s = (3/2)\gamma^2\epsilon$, implying $\epsilon\epsilon_s = 3/32$, i.e., $m_ec^2 \epsilon_s = 2.4$ GeV, consistent with the measured break energy. Spectral calculations employing a power-law electron spectrum produce a break at the right energy, but with an insufficiently sharp break \cite{2010ApJ...721.1383A}. Combined Compton and synchrotron losses will harden the electron spectrum and reduce the magnitude of the break \cite{2002ApJ...568L..81D,2003A&A...406..855M}, but a curved electron spectrum described by a log parabola function \cite{2004A&A...413..489M}, formed when acceleration dominates over cooling, might resolve this difficulty.

\subsection{Blazar sequence}

\begin{figure}[h]
\includegraphics[width=24pc]{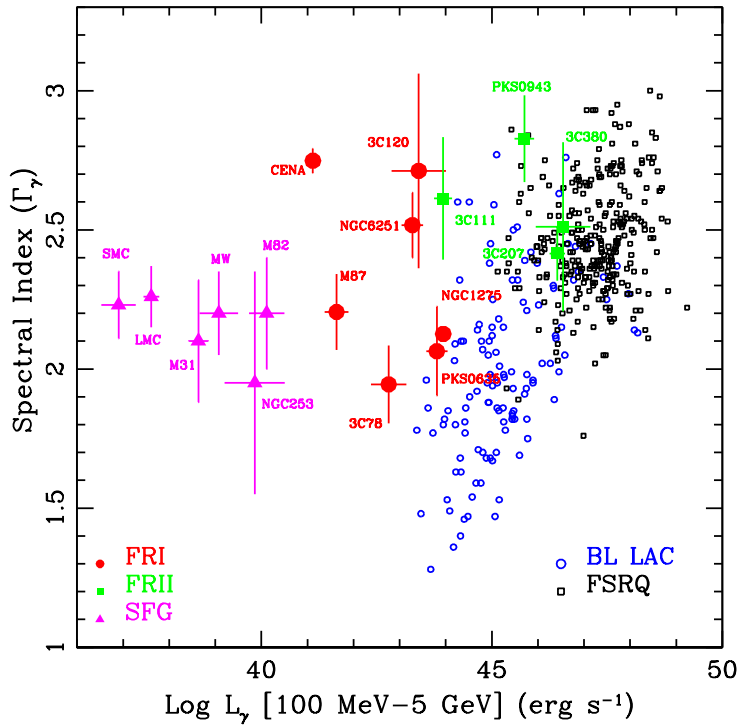}\hspace{2pc}%
\begin{minipage}[b]{12pc}
\caption{\label{GammavsL}
Gamma-ray spectral slopes $\Gamma_\gamma$ of BL Lac objects (open blue circles),  FSRQs (open black squares), FR1 radio galaxies (red circles), FR2 radio sources (green squares), and star-forming galaxies (magenta diamonds), are plotted as a function of their 100 MeV -– 5 GeV $\gamma$-ray luminosity \cite{2010ApJ...720..912A}. The $\gamma$-ray emitting misaligned AGNs are 1LAC sources associated with objects in the Third Cambridge Revised (3CRR) and Molonglo radio catalogs. The $\gamma$-ray emission from star-forming galaxies are powered by cosmic-ray processes rather than AGNs.}
\end{minipage}
\end{figure}

The search for an organizing principle in blazar studies has centered on the so-called blazar sequence, which is the well-known anti-correlation of the $\nu F_\nu$ synchrotron and Compton peak frequencies with luminosity \cite{1998MNRAS.299..433F,1998MNRAS.301..451G}. It is reflected in the Fermi data by plotting the blazar spectral index  $\Gamma_\gamma$ versus the $\gamma$-ray luminosity. This figure reveals a clear separation between BL Lac objects and FSRQs---a blazar divide  \cite{2009MNRAS.396L.105G}---that is attributed to a difference in accretion regime.  As plotted in Figure \ref{GammavsL}, one sees that all FSRQs have $\Gamma_\gamma > 2$. Inasmuch as the presence of strong emission lines means that there has to be a significant surrounding scattered radiation field, then a consistent interpretation of  FSRQ SEDs is to attribute the $\gamma$-ray emission to external Compton processes  \cite{1998MNRAS.301..451G}, with the trend in the SEDs from FSRQ to BL Lac objects resulting from the reduction in the intensity of the surrounding radiation field \cite{2002ApJ...564...86B}. This is reasonable provided that the emission region of FSRQs is found within the BLR. Indeed, this seems to be the most natural explanation for rapid variability \cite{2011A&A...530A..77F}. Yet this expectation collides with detection of VHE $\gamma$ rays from FSRQs like 3C 279 \cite{2008Sci...320.1752M} and, most recently and most spectacularly, 4C +21.35, which exhibits a 10 minute flaring episode of 70 -- 400 GeV radiation \cite{2011ApJ...730L...8A} from a source with very strong IR radiation \cite{2011ApJ...732..116M} that would strongly attenuate high-energy radiation made within the pc scale. Two-zone models have been advocated to deal with these questions \cite{2009ApJ...703.1168B,2011arXiv1104.0048T}. 

The blazar sequence and blazar divide are not, however, immune to criticisms, most notably that a large fraction of BL Lac objects lack redshifts. This fraction amounts to more than 50\% in the 2LAC. Because strongly beamed nonthermal radiation would conceal emission lines, then the highest redshift, highest luminosity BL Lac objects would be selectively excluded from the blazar sequence diagram, and these would also likely be the HSP BL Lacs where the nonthermal radiation is strongest in the optical/UV range. Besides outliers to the blazar sequence, there is evidence for a population of HSP FSRQs \cite{2003ApJ...588..128P}. A further concern is that the BL Lac objects are primarily identified from and biased towards the radio and X-ray surveys from which they are selected \cite{2007Ap&SS.309...63P}. 

In any case, the simplest characterization of the blazar sequence must consider the contribution of slightly off-axis jetted sources and misaligned radio galaxies. When observing away from the beaming axis, the apparent luminosities from the smaller Doppler boosting should appear at lower peak frequencies, opposite to the blazar sequence correlation. Consequently radio galaxies like NGC 6251 \cite{2011arXiv1107.4302M} will fill in a distinct portion of a diagram of the blazar sequence, turning it rather into an envelope \cite{2011arXiv1107.5105M}.

\subsection{Relation between aligned and misaligned jet sources}

A curiosity revealed by Fig.\ \ref{GammavsL} has to do with the spread in luminosities of the radio galaxies compared with the aligned blazar jet sources. The misaligned AGN (MAGN) population detected with Fermi consists of 11 sources  \cite{2010ApJ...720..912A}. Theses includes 7 FRI and 2 FRII radio galaxies, which are thought to be the parent population of BL Lac objects and FSRQs, respectively, in addition to 2 steep spectrum radio quasars that are believed to be slightly misaligned FSRQs. The MAGN sources in the 3CRR catalog have large core dominance parameters compared to the general 3CRR source population, implying that the beamed component makes an appreciable contribution to the $\gamma$-ray flux. This is furthermore supported by the fact that three of these sources---3C 78, 3C 111, and 3C 120---do not appear in 2LAC, evidently due to variability. 

Why the ratio of measured $\gamma$-ray luminosities of FRI galaxies and BL Lac objects span a much larger range than the comparable ratio for FRII galaxies and FSRQs is an open question, if in fact it is not due to statistics. The simplest possibility is that there are a lot more nearby FRIs, and limitations on Fermi sensitivity will therefore favor detection of these nearby sources. At $z\ll 1$, the luminosity distance $d_L\approx 4200z(1+z)$ Mpc. The {\it Fermi}-LAT reaches a limiting energy-flux sensitivity of $\approx 5\times 10^{-12}$ erg cm$^{-2}$ s$^{-1}$ for two years of observations that is, unlike integral photon flux, only weakly dependent on source spectral index \cite{2LAC}.  Fermi can thus only detect sources with luminosity $L_\gamma \gtrsim 10^{46}z^2(1+2z)$ erg s$^{-1}$. The number of nearby FRI and FRII galaxies depends on the space density of these objects which, in the study by Gendre, Best, and Wall \cite{2010MNRAS.404.1719G}, is inferred from 1.4 GHz NVSS-FIRST radio observations (recall that the Fanaroff-Riley dichotomy is based on 178 MHz luminosities $P_{178~{\rm MHz}}$). From their Figure 12,  FRI and FRII radio galaxies have a local ($z<0.3$) space density of $\approx 35\times 10^{-6}$ Mpc$^{-3}$ and $\approx 2\times 10^{-6}$ Mpc$^{-3}$, respectively, at $P_{1.4~{\rm GHz}} \gtrsim 10^{22}$ W Hz$^{-1}$ sr$^{-1}$, implying that the local space density of FRIs exceeds that of FRIIs by $\approx 20$.

This implies that there are several hundred FRIs and FRIIs within $z \cong 0.1$, yet other than 3C 111, all LAT-detected radio galaxies within this volume are FRI sources (see Figure 3 of \cite{2010ApJ...724.1366D}).  This is further remarkable in that FSRQs and their putative parent population of FRII radio galaxies are far more radio luminous than BL Lac objects and FRI galaxies.  With only a few hundred randomly aligned sources within $z = 0.1$, it might be the case that a narrower $\gamma$-ray beaming cone in FSRQs, with a more rapid fall-off in off-axis flux, makes detection of these nearby sources far less likely than a broader $\gamma$-ray emission cone in BL Lac objects, as expected for the different beaming factors from SSC emission and external Compton processes \cite{1995ApJ...446L..63D,2001ApJ...561..111G}.  This could also reflect differences in jet structure between FSRQs and BL Lac objects \cite{2000A&A...358..104C}, or extended jet or lobe emission in FRIs that is missing in FRII galaxies.

\subsection{Apparent super-Eddington luminosities of blazars}

We end this contribution on a speculative note. The time-averaged luminosities of FSRQs  extend to values in excess of $L_\gamma\approx 10^{49}$ erg s$^{-1}$ (Figure \ref{GammavsL}). In the extraordinary 2010 November flare of 3C 454.3 \cite{2011ApJ...733L..26A}, $L_\gamma$ reached apparent isotropic luminosity of $(2.1\pm 0.2)\times 10^{50}$ erg s$^{-1}$ over a period of a few hours, making it the most luminous blazar yet observed. Black-hole mass estimates for 3C~454.3 are in the range $0.5 \lesssim M_9 \lesssim 4$ \cite{2011MNRAS.410..368B}, where $10^9 M_9M_\odot$ is the mass of the black hole powering this AGN. For this range of masses, the Eddington luminosity therefore  ranges from $\approx 6\times 10^{46}$ erg s$^{-1}$ to $\approx 5\times 10^{47}$ erg s$^{-1}$. During this extreme outburst, the apparent luminosity of 3C 454.3 was more than a factor of $\approx 400$ greater than its Eddington luminosity. Even its time-averaged luminosity of $L_\gamma\approx 10^{49}$ erg s$^{-1}$ is super-Eddington by a factor of $\approx 20$. 

Assuming that the Eddington condition {\it does} limit accretion flow onto the black hole, which is likely to be the case for the long-term average luminosity if not for the flaring luminosity, then the absolute radiant luminosity is limited to a value of $L_{abs} \lesssim  5\times 10^{47}$ erg s$^{-1}$. This is consistent with the large apparent luminosities if the emission is highly beamed. For a simple top-hat jet beaming factor, a jet opening angle $\theta_{j}$ implies a beaming factor $f_b = \theta_j^2/2$ for a two-sided jet with $\theta_j \ll 1$. A mechanism for collimation is, however, required. Should this arise from the Blandford-Znajek process, then we are still restricted to values of the absolute Blandford-Znajek power
\begin{equation}
{d{\cal E}\over dt}|_{\rm BZ} \approx \pi c ({a\over M})^2 r_+^2 B_+^{r 2}\;,
\label{PBZ}
\end{equation}
where $a/M$ is the angular momentum per unit mass, $r_+$ is the radius of the event horizon, and $B_+^r$ is the radial component of the magnetic field threading the event horizon \cite{2006tbhr.book..119L}. Scaling the energy density of the magnetic field to the energy density of accreted matter near the event horizon shows that the Blandford-Znajek power is likewise Eddington-limited. Making the hypothesis that the extraction of energy through black-hole rotation collimates the jet outflow with $\cos\theta_j \approx a/M$, then $f_b \cong 1 - (a/M)$ and $\theta_j \cong \sqrt{2(1-a/M)}$, implying $a/M > 1 - (L_{\rm Edd}/L_\gamma)$. If the jet opening angle is a consequence of the bulk Lorentz factor $\Gamma$ of the outflow, then $\Gamma \gtrsim \sqrt{L_\gamma/2L_{\rm Edd}}$.

For the case of 3C 454.3 in its flaring state, when $L_\gamma/L_{\rm Edd}\gtrsim 10^3$, this hypothesis then implies that $a/M > 0.999$ and $\Gamma \gtrsim 23$, which is consistent with the value $\Gamma_{min} \approx 14$ from $\gamma\gamma$ opacity arguments \cite{2011ApJ...733L..26A}. In the most conservative case with $M_9 = 4$, $a/M \cong 0.998$ and $\Gamma \gtrsim 15$. This is  marginally consistent with the limiting maximum value $a/M \cong 0.996$ suggested by Aschenbach from analyses of microquasars and the Galactic Center black hole \cite{2004A&A...425.1075A}. The smaller black-hole mass estimate,   $M_9 = 0.5$ (Bonnoli et al. 2010), implies a value of $a/M$ that violates this limit by a large margin.  Much work, both numerical and theoretical, has been devoted to jet formation from the Blandford-Znajek process, and it is unclear if jet collimation can be described by the guess that  $\cos\theta_j \approx a/M$, but it is interesting to suggest this possibility, which leads to values of $\Gamma$ consistent with separate inferences regarding the outflow Lorentz factor.

\ack
This work is supported by the Office of Naval Research and the Fermi Guest Investigator program. We thank Paola Grandi for preparing Fig.\ \ref{GammavsL} and stressing the unusual properties of LAT-detected radio galaxies, and Maxim Barkov, Amir Levinson, Govind Menon, Juri Poutanen, Anita Reimer, and Boris Stern for discussions. CD would also like to thank the organizers, especially Robert Wagner, for the opportunity to visit Finland.

\section*{References}

\bibliography{Dermer_Lott}

\end{document}